# TESTING A QUANTUM COMPUTER

JACOB D. BIAMONTE[1] AND MAREK A. PERKOWSKI [1, 2]

[1] *Quantum Information Laboratory*
*Department of Electrical Engineering, Portland State University,*
*Portland, Oregon, 97207-0751, USA*

[2]*Department of Electrical and Computer Engineering, Korea Advanced Institute of*
*Science and Technology, Taejon 305-701, Korea*

Corresponding author: biamonte@ieee.org

*Abstract*: We address the problem of quantum test set generation using measurement from a single basis and the single fault model. Experimental physicists currently test quantum circuits exhaustively, meaning that each *n*-bit permutative circuit requires $\zeta \times 2^n$ tests to assure functionality, and for an *m* stage permutative circuit proven not to function properly the current method requires $\zeta \times 2^n \times m$ tests as the upper bound for fault localization, where *zeta* varies with physical implementation. Indeed, the exhaustive methods complexity grows exponentially with the number of qubits, proportionally to the number of stages in a quantum circuit and directly with *zeta*. This testability bound grows still exponentially with the attempted verification of quantum effects, such as the emission of a quantum source. The exhaustive method will soon not be feasible for practical application provided the number of qubits increases even a small number from the current state of the art.

An algorithm is presented making fault detection feasible both now and in the foreseeable future for quantum circuits. The presented method attempts the quantum role of classical test generation and test set reduction methods known from standard binary and analog circuits. The quantum fault table is introduced, and the test generation method explained, we show that all faults can be detected that impact calculations from the computational basis. It is believed that this fundamental research will lead to the simplification of testing for commercial quantum computers.

## 1. INTRODUCTION

Since its inception, the microelectronics industry has progressed by shrinking circuitry [3]. Moore's Law [4] can be interpreted as the rate of mankind's advancement towards making commercial electronic devices on the atomic scale. Based on the current rate CMOS technology is progressing, it is predicted that large scale designs at the level of a single atom will be very possible in the next 20 years. The reduction of circuits to the atomic scale not only brings with it the idea of reversibility, leading to virtually energy free computation as shown by Landauer [5], and Bennett [6], but that of quantum effects. Quantum Information theory represents the known physical limitations our universe places on man's ability to construct information processing machines. In short, the quest for understanding physics is now intertwined with mankind's ability to build exponentially faster computers [28].

A major problem faced by physicists when attempting to create a practical quantum computer are the natural imperfections inherent in any quantum system. Currently experimental physicists have only begun to experience a need to research optimized testing methods due to the small qubit count of current quantum circuits. In addition, the slow rate of progress at physically realizing certain quantum circuits has made the idea of rapid testing less feasible. For example in NMR it can sometimes take months to fine tune the sequence of pulses necessary to implement a simple universal gate that functions properly. The current approach is to compare all inputs with all outputs many times as means of verification for a given circuit. This amounts to $\zeta \times 2^n$ distinct tests for permutative circuits and is known in classical logic as the exhaustive method [13]. When diagnostic methods of fault localization are needed, the physicist typically uses a property referred to as perfect observability [14] combined with the exhaustive method, where each stage of the circuit is probed by measurement. As such an *m* stage permutative circuit has an upper testability bound of $\zeta \times 2^n \times m$ distinct tests, where *zeta* represents iterative testing intended to assure functionality which varies depending on physical implementation. Observe, as will be shown below, that this method does not always give 100% assurance of circuit correctness, as a qubit can for example become coupled with that expected from measurement and a state of erroneous nature. This preliminary work introduces only the concept of quantum circuit verification, the reader interested in quantum fault diagnostics such as fault localization should refer to [15, 16].

## 2. BACKGROUND

In classical circuit design an error in a circuit is typically referred to as a fault – in this paper we use the word error and fault interchangeably when referring to both classical and quantum circuits. In classical circuits we represent the occurrence of faults digitally as the inversion of one or more bits of information at one or more locations in the circuit [14, 17, 18]. In analog circuits a fault is manifested by lost signal integrity at one or more stages in the circuit. The detection of classical faults broadly falls in two distinct types of tests, the first being parametric tests, as those typically adapted by an analog test engineer, taking into account parametric measures [14], and the second is called Logical test [14], (*known also as functional test* [14, 17, 18, 19]) in which the functional output of a system is compared with the expected output value for a given input [14, 17, 19].

In a quantum circuit a fault is said to be observed if one or more bits have a different value from that anticipated as the logical output. In this work, we concern ourselves with logical testing as applied to quantum circuits, where we inspect the logical data processed by the quantum circuit and compare this data with the expected values allowing us to make a judgment on functionality.

To further narrow the category of logical testing, we now form three categories of fault occurrence where both classical and quantum circuits suffer. It is common to have faults inherent in the circuit by manufacturing errors (*system faults* [14], *Manufacturing faults* [14], *Physical malfunctions/defects* [14, 17, 19]), faults introduced by the programmer (*design errors* [14], *Conceptual Errors* [14], *programming errors/bugs*) and random errors caused by the environment (*transient failures* [14], *soft errors* [18,19], *random errors* [2], *probabilistic faults*).

As noted, similarity exists between the categories of classical circuit faults and quantum circuit faults, with several very notable exceptions. The first is that a quantum circuit may have a fault always present that is never detectable when measured with a given projector, and the second is that a quantum circuit may always have a fault present that is only detectable in some percentage of measurements from any basis. Generally this amounts to the condition of a phase shift, where a relative phase shift can be detected while overall phase shifts cannot. This gives rise to a new definition needed to represent faults with these unique properties for the classical test engineer adapting methods to quantum circuits. In classical circuits the phrase "<u>probabilistic fault</u>" (*with* *contributions from* [14, 17, 19]) and the phrase "deterministic fault" (*with contributions from* [14, 17, 19]), should be used with caution when referring to a fault in a quantum circuit, this terminology is misleading to those already familiar with classical testing, as they take different meanings in that field. To avoid confusion the faults defined here will be coined "quantum faults", as is the case for the remainder of this paper. Although a quantum fault can occur probabilistically, such as that introduced by environmental noise, our method deals with troubleshooting and detection, and is useful for verification of repeatable faults (*although it can be adapted and used statistically*), this difference is made more distinct by referring now to several works on error correcting codes meant to neutralize random errors [22, 23, 24, 27, 26]. A practical quantum computer should be equipped with an error correction mechanism and will need verification if correct functionality is stated; it is this verification that relies on quantum test, or more generally state tomography.

Although it may seem as if testing quantum circuits is similar to testing reversible circuits, there is actually little specific similarity between the two. An in-depth study of quantum circuits leads one to understand that although quantum circuits are by definition reversible, the fundamental differences between the inner workings of quantum gates and the gates used in other implementations of reversible logic have a drastic impact on testing. For example, a study recently done in [2] noted that, "*Each test vector covers exactly half of the possible faults, and each fault is covered by exactly half of the possible test vectors*." However, we found a similar property of symmetry not to exist in quantum circuits. On the other hand, the theories used in testing classical logic circuits, especially in [14, 17, 19, 25], form the foundation for the new types of theories developed and presented here.

## 3. THE FAULTS OF EVOLUTION

There exist three main categories of faults in a quantum computer that we consider: faults in preparation, faults in evolution and faults in measurement. In this paper as means of simplification evolutionary faults are assumed exclusively. In other words this preliminary work assumes what is referred to in binary testing as the single fault model [2, 14, 17, 19]. One can now judge the case where an evolutionary error is unique, such that the preparation and measurement operations are ideal.

The Hamiltonian of the spin system that models Ising type interactions is given by,

$$\hat{H} = \sum_{i,\alpha} a_{i\alpha} \sigma_{i\alpha} + \sum_{i,j} J_{ij} \sigma_{iz} \sigma_{jz} \quad \text{\textit{Equation 1 – The spin Hamiltonian}}$$

Let us assume the Hamiltonian takes a form such as $\hat{H}_\varepsilon = \hat{H} + v$, with $v$ representing a small error. The resulting impact in the presence of this error manifests in the form of changing the probability amplitudes of the possible outcomes, such that there is an altered chance for a particular outcome in measurement resulting from the presence of $v$ [21]. We state that an error takes one of two forms, the first being the presence of an error and the subsequent being the lack of an error. The detection of an error is dependent on the choice of a measurement basis and the chance of detection, but the presence of an error is not, giving rise to the risk of confusion and that defined as a "quantum fault."

Suppose we wish to examine $v$ in further detail. From [29] one can study a CNOT gate accurate to phase.

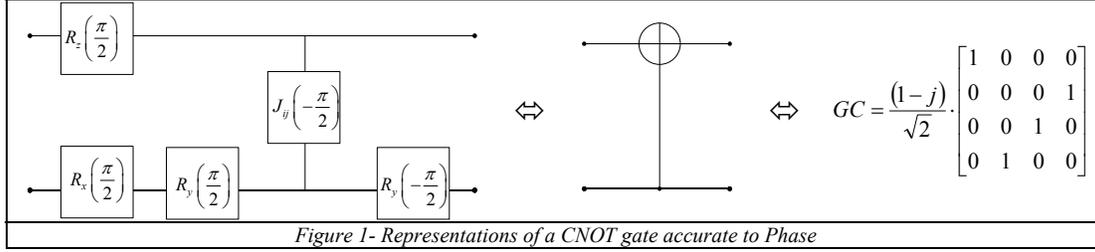

*Figure 1- Representations of a CNOT gate accurate to Phase*

Also from [29] the alternate representation of Figure 1 is presented in a form implemented with the presented spin Hamiltonian

$$CNOT_{ij} = R_{iz}\left(\frac{\pi}{2}\right) R_{jx}\left(\frac{\pi}{2}\right) R_{jy}\left(\frac{\pi}{2}\right) J_{ij}\left(-\frac{\pi}{2}\right) R_{jy}\left(\frac{-\pi}{2}\right).$$

*Equation 2 – Possible sequence of operations needed for implementation of the CNOT gate*

In this work $v$ denotes the addition of an error in terms of Equation 1, with this addition occurring in the time sequence needed to implement a quantum information process. However it is not clear if $v$ is predestined to be that of a unitary fault, as the system in question is a component of a larger Hilbert space.

We take now an example such that the situation of an error may reside before or after any element of the pulse sequence in Equation 2. Even the removal of a gate is modeled this way, such that arbitrary gate $G$ has a nearest neighbor as its conjugate transpose $G^\dagger$, where $GG^\dagger = G^\dagger G = I$. In practice defining the correct general fault model remains a technology specific issue, however the removal of a gate should actually be brought with the insertion of another since a machine is reasonably more likely to perform an erroneous operation than to miss it completely.

## 4. QUANTUM FAULT MODEL

Although unwanted interaction in a quantum system will be present all the time, the most widely cited error correcting codes neglect interaction among qubits focusing on single qubit rotations defined by the Pauli group. This work only outlines a broad method that can be used in quantum test. Although the fault model presented here may prove useful for the rapid verification of commercial quantum devices, specific fault models must be used to verify current technology. The quest for a viable general fault model is not the goal of this paper; here we restrict ourselves to the problem of test generation and not implementation specific fault models. It is the intent of the presented method to be so general that adaptation to alternate technology is just a matter of defining a new fault model.

Traditionally the Test Generation Problem is thought of as the generation of a sequence of tests, *(test set)* that when applied to a circuit and compared with the circuit's output, will either verify that the circuit is correct or will determine that it contains one or more faults [14]. In other words, testing is the verification of functionality, and running the ideal test set amounts to complete system verification. For the single fault model, it is the typical case that certain single tests can verify the existence of a fault at multiple locations in a circuit at the same time, thus it is the goal to choose the fewest tests possible needed to determine all possible errors.

Based on the ideal model of a quantum computer adapted from [29] and that aforementioned for classical binary circuits we can now make certain assumptions about the nature of faults present in a quantum circuit. For our work we will restrict ourselves to the quantum fault model of inserting any gate from the single qubit operations defined by the Pauli group (*which is the fault model assumed in error correcting codes*), with the addition to the removal of any gate that is represented in the quantum circuit schematic. As can be seen in Figure 2 we have defined the locations that errors are thought to reside for the preparation *(P1, P2)*, evolution *(E1-E8)* and measurement *(M1, M2)* stages of the circuit.

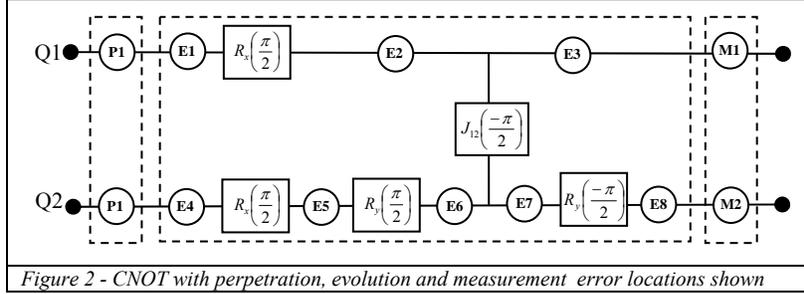

*Figure 2 - CNOT with perpetration, evolution and measurement error locations shown*

Formally we define our fault model as follows: The Pauli group on *n*-qubits is a subgroup of linear operators on $W := (C^2)^{\otimes n}$ defined as,

$$G^n = \{\pm I, \pm X, \pm Y, \pm Z, \pm jI, \pm jX, \pm jY, \pm jZ\}^{\otimes n}$$

and in this work we restrict ourselves for simplicity to the set of generators defined as,

$$X = \begin{bmatrix} 0 & 1 \\ 1 & 0 \end{bmatrix}, Y = \begin{bmatrix} 0 & -j \\ j & 0 \end{bmatrix}, Z = \begin{bmatrix} 1 & 0 \\ 0 & -1 \end{bmatrix}, I = \begin{bmatrix} 1 & 0 \\ 0 & 1 \end{bmatrix}$$

with the addition of gate removal this defines our quantum fault model. In the future perhaps a new fault model outside of the Pauli group may be a faulty interaction gate arising in implementation as a result of faults in the interaction of qubits. This will be an area of the authors' further study.

## 5. THE EXISTENCE OF COMPLETE TEST SETS

A goal of quantum test is to find the optimum test set that detects all possible faults that can be present in the system with the least number of test vectors; this is known as high defect coverage for both classical and quantum circuits [14]. In general when we perform a given test, we are attempting to determine to a certain degree of assurance whether or not a fault is present. For example, let us assume that we have defined a location in a theoretical quantum circuit as location **X**.

If we insert a fault at location **X** and perform a calculation telling us that the fault we inserted amounts to a bit flip when measured, then if we measured an actual circuit and find the binary output has been inverted *(based on our model)*, we can say that we have determined that the fault exists at **X** *(we have detected and in this case localized X)*. However, even in this simple case one can make no statements as to the probability that a fault is present at location **X** in the circuit prior to running a given test. It is only running a single test that allows us to make one of the following statements: (*1*) *the fault is definitely not present at location* **X***, or* (*2*) *the fault is definitely present at location* **X**.

Now if we perform the same calculations for a fault at location **X**, and this time determine that the fault we inserted only occurs with a given probability denoted by **P(x)**, prior to running any single test we can make no statement related to the occurrence of **P(x)**. However, after running any number of tests we can say one of two things: *(1) the fault is definitely present at location* **X**, *or (2) we are able to judge the probability of the assumption of the occurrence of the fault at location* **X** *based on accepted experimental result.* In other words, we are only able to judge our assumption to a given proportion, and this postulation never answers the question of whether or not a fault is strictly present based on a given fault model for many quantum circuits. We base these results on the formulation of the quantum covering problem presented in this work.

## 6. PREVIOUS WORK ON TESTING BINARY CIRCUITS

In order to illustrate our method of fault detection for quantum circuits we will illustrate the method used in [2] which uses a direct approach to generate a test set that will detect all faults in a binary reversible logic circuit by decomposing larger circuits into smaller sub-circuits (*block partitioning*). The problem of finding a minimal test set is solved using Integer Linear Programming (ILP). The authors of [2] use the single stuck-at fault model [14, 25, 17, 36] to detect faults in internal lines and primary input and output lines of the circuit. Their main contributions are the following observations regarding reversible circuits: **(1)** Any test set that is complete for the single stuck-at fault model is also complete for the multiple stuck-at fault model. **(2)** Each test vector covers exactly half of the faults, and each fault is covered by exactly half of the possible test vectors.

However, the best way to illustrate the approach of simply verifying a circuit's functionality is by introducing the circuit in Figure 3 whose corresponding truth table is shown in Table 1.

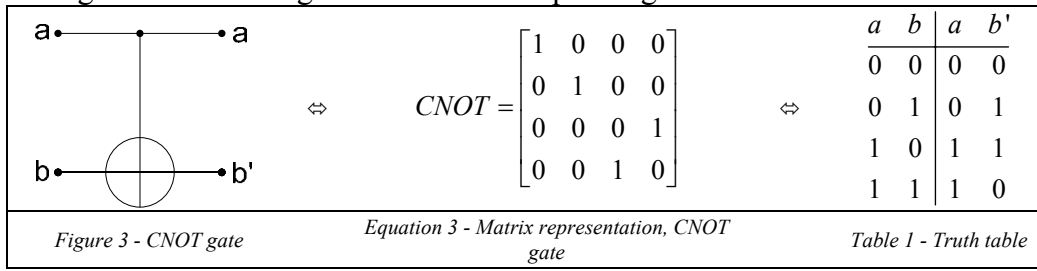

*Figure 3 - CNOT gate*    *Equation 3 - Matrix representation, CNOT gate*    *Table 1 - Truth table*

The preparatory steps in fault detection is to examine the circuit in question and label all locations of which a fault is thought to possibly reside, as done in Figure 4. When this is done a fault model must be selected, for this example we use the following,

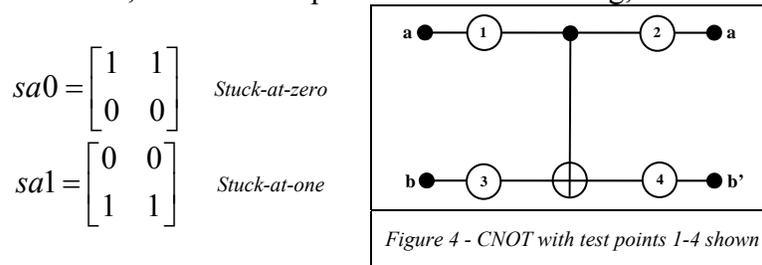

*Figure 4 - CNOT with test points 1-4 shown*

For each location depicted as one that could possibly be that containing a fault, each fault model is inserted and the circuits truth table is again recalculated, as shown for a particular case in Figure 5.

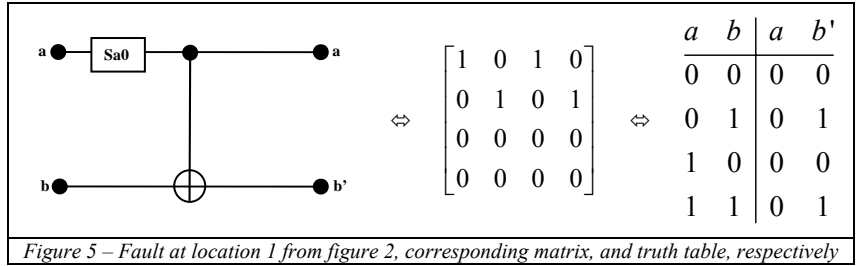

*Figure 5 – Fault at location 1 from figure 2, corresponding matrix, and truth table, respectively*

After each iteration another column of a fault table is created. For our purposes we define a fault table as that having all tests as rows and all faults as columns, with the first column representing that of a good circuit. Typically it is the case that a "1" at the intersection of row $R_i$ and column $C_j$ determines that test $R_i$ detects fault $C_j$ for the case of fault detection, but in Table 2 we show this intersection to mean the output of the circuit, with that of a detectable fault depicted in bold faced type (*we named this a reversible fault table*). In Table 3 we proceed to introduce what will be called a *classical fault table*. In this table, rows with entries of 1 mean that a probability of 1 exists in covering columns for a test depicted by a given row.

| a b | GC | Sa0@1 | Sa0@2 | Sa0@3 | Sa0@4 | Sa1@1 | Sa1@2 | Sa1@3 | Sa1@4 |
|-----|----|-------|-------|-------|-------|-------|-------|-------|-------|
| 00  | 00 | 00    | 00    | 00    | 00    | **11**| **10**| **01**| **01**|
| 01  | 01 | 01    | 01    | **00**| **00**| **10**| **11**| 01    | 01    |
| 10  | 11 | **00**| **00**| 11    | **10**| 11    | 11    | **10**| 11    |
| 11  | 10 | **01**| **00**| 11    | 10    | 10    | 10    | 10    | **11**|

*Table 2 - Fault table for CNOT gate*

| a b | GC | Sa0@1 | Sa0@2 | Sa0@3 | Sa0@4 | Sa1@1 | Sa1@2 | Sa1@3 | Sa1@4 |
|-----|----|-------|-------|-------|-------|-------|-------|-------|-------|
| 00  | 00 | 0     | 0     | 0     | 0     | **1** | **1** | **1** | **1** |
| 01  | 01 | 0     | 0     | **1** | **1** | **1** | **1** | 0     | 0     |
| 10  | 11 | **1** | **1** | 0     | **1** | 0     | 0     | **1** | 0     |
| 11  | 10 | **1** | **1** | 1     | 0     | 0     | 0     | 0     | **1** |

*Table 3 - 1's represent detectable faults*

We will use the greedy approach of picking the row with the highest depict coverage, and as noted in [2], it makes little difference in this case, as we chose our first test vector to be |00>. We now repeat this for Table 4, and Table 5.

| a b | GC | Sa0@1 | Sa0@2 | Sa0@3 | Sa0@4 |
|-----|----|-------|-------|-------|-------|
| 01  | 01 | 0     | 0     | **1** | **1** |
| 10  | 11 | 1     | 1     | 0     | 1     |
| **11** | 10 | 1  | 1     | 1     | 0     |

*Table 4 – From Table 3, removal of the row '00', after running test vector |00>*

| a b | GC | Sa0@4 |
|-----|----|-------|
| 01  | 01 | **1** |
| **10** | 11 | **1** |

*Table 5 – From Table 4, removal of the row '11', after running test vector |11>*

It is now clear that the test set: {00, 11, 10} is complete for our fault model, as shown in Figure 6. In other words, this test set will detect all possible faults in the circuit provided all faults are of the type specified by the fault model.

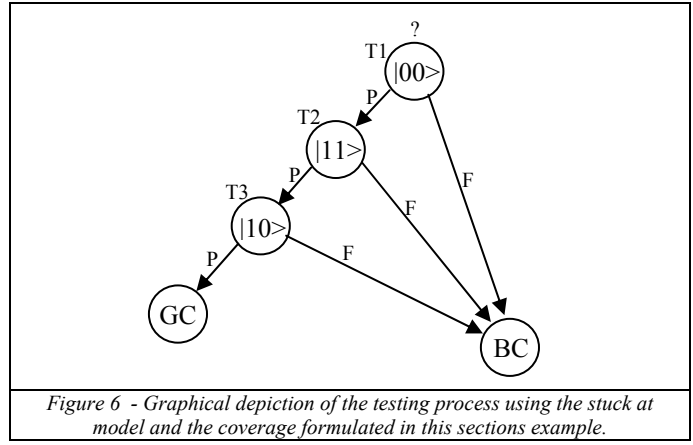
*Figure 6 - Graphical depiction of the testing process using the stuck at model and the coverage formulated in this sections example.*

## 7. THE QUANTUM COVERING PROBLEM

In [37] the notion of probabilistic set-covering is described as the generation of a random binary vector and the covering constraint has to be satisfied with some prearranged probability. Traditionally we represent $\eta$ as an element in the set to be covered where $\eta$ is constrained such that $\eta \in \{0,1\}^n$. A distinction is in order defining an entry in the quantum fault table depicted by $\eta$, where $\eta$ is constrained such that $0 \leq \eta \leq 1$, as such it is clear that the traditional covering problem is a special case where $\eta$ takes the extreme values and that defined is a more general formalism. With this mentioned the notion of probabilistic set-covering as defined in [37] does not apply. Furthermore, the problem defined herein is only solvable if a constraint is defined for columns that contain no elements of additive identity and multiplicative unity. In other words, our problem is now formulated differing from the classical case with the addition of positive fractional entries, arising from for example the interaction of qubits. This addition makes the concept of full cover one that is not achievable for testing some quantum circuits. Let us assume the existence of only a single fault, given the table

| $\ddots$ | $f_a$ | $f_{a+1}$ | $f_{a+2}$ | $\cdots$ | $f_n$ |
|---|---|---|---|---|---|
| $T_a$ | $P(a_1)$ | $P(a_2)$ | $P(a_3)$ | $\cdots$ | $P(a_n)$ |
| $T_{a+1}$ | $P(b_1)$ | $P(b_2)$ | $P(b_3)$ | $\cdots$ | $P(b_n)$ |
| $T_{a+2}$ | $P(c_1)$ | $P(c_2)$ | $P(c_3)$ | $\cdots$ | $P(c_n)$ |
| $\vdots$ | $\vdots$ | $\vdots$ | $\vdots$ | $\ddots$ | $\vdots$ |
| $T_n$ | $P(n_1)$ | $P(n_2)$ | $P(n_3)$ | $\cdots$ | $P(n_n)$ |

if we sample over arbitrary space $\{T_a, T_{a+1}, \cdots, T_n\}$, taking into account the disjoint cover of arbitrary $f_a$, we can formally denote

$$\Gamma = P(a_1) + P(\bar{a}_1 \cap b_1) + P(\bar{a}_1 \cap \bar{b}_1 \cap c_1) + \cdots + P(\bar{a}_1 \cap \bar{b}_1 \cap \bar{c}_1 \cap \cdots \cap n_n) \leq 1$$

as the space covered by the test set $\{T_a, \cdots T_n\}$, where $P(\bar{x}) = 1 - P(x)$. As further example illustrating a particular point in sampling over arbitrary space $T_a$ with the number of samples meaning *n* times for arbitrary $f_a$ the cover is given by

$$\Gamma = P(a_1)^1 + P(a_1)^2 + P(a_1)^3 + \cdots + P(a_1)^n = \sum_{n=1}^{n} P(a_1)^n$$
$$= P(a_1) + P(\bar{a}_1 \cap a_1) + P(\bar{a}_1 \cap \bar{a}_1 \cap a_1) + P(\bar{a}_1 \cap \bar{a}_1 \cap \bar{a}_1 \cap \cdots \cap a_n) \leq 1.$$

Thus we define the *cover* of the quantum fault table for arbitrary test order $\{T_a, T_{a+1}, \cdots, T_n\}$ as $\Gamma$ for arbitrary column $f_a$. Formally we state that the quantum covering problem reduces to full cover to only a margin of assuredness except in the case of infinite iterative testing, given that each entry in the column of a fault table is less than one with an entry greater than zero.

Let us further illustrate this point by examining the following example. For the contrived fault table,

|       | $f_1$ | $f_2$ | $f_3$ |
|-------|-------|-------|-------|
| $T_1$ | .9    | .1    | .6    |
| $T_2$ | .6    | .8    | 1     |
| $T_3$ | .2    | .9    | .7    |

let us assume we sample over the space $\{T_1, T_2, T_3\}$. The cover is now calculated for each of the columns,

$$C_1 = .9 + (1-.9)(.6) + (1-.9)(1-.6)(.2)$$
$$C_2 = .1 + (1-.1)(.8) + (1-.1)(1-.8)(.9)$$
$$C_3 = .6 + (1-.6)(1) + (1-.6)(1-1)(.7)$$

## 8. QUANTUM FAULT DETECTION EXAMPLE

The pulse sequence presented in Equation 2 is of interest here. As explanation the circuit schematic from Figure 2 is again redrawn with the addition of several new labels that will become apparently necessary in the explanation of the presented algorithm. The first labels are that of the first and second qubits in the system denoted as Q1, and Q2 respectively. The circuit is again broken up into smaller portions such that a stage is defined to mean each individual pulse labeled here as S1-S5. Each pulse has the dual presence of a nearest neighbor referred to here as a division and labeled D1 through D6. It is important to note that errors in the circuit model occur in time, as time passes from left to right in our quantum circuit schematic. It is worthwhile for some readers if we mention that multiple stages in this circuit could have been done in one pulse, and the circuit may not be minimal, however, this is implementation specific, and are problem is formulated around the Hamiltonian from Equation 1. It is hoped that the reader will immediately make adaptations to the presented algorithm, as the goal of this work is to present an idea in the clearest possible manor, and many simplifications were made, such as the use of a single fault model and measurement from a single basis.

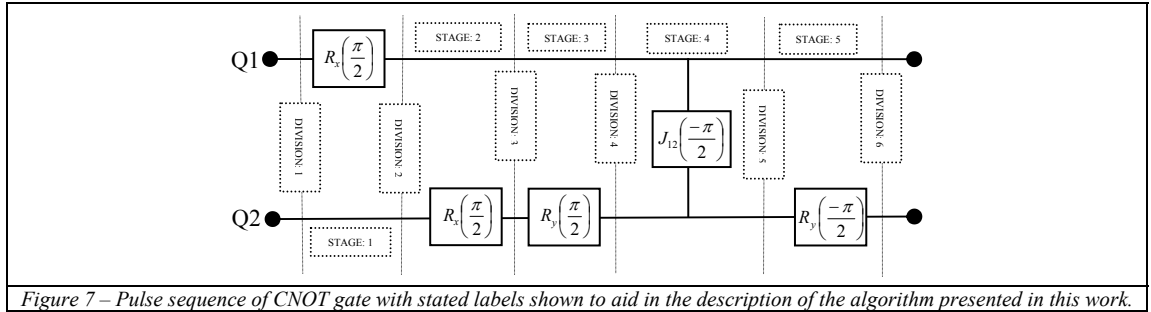
*Figure 7 – Pulse sequence of CNOT gate with stated labels shown to aid in the description of the algorithm presented in this work.*

Under the assumption of the single fault model we can clearly define the faults used in this work as any of the first three Pauli (X, Y, or Z) rotations acting on a single qubit or the addition of any single pulse (S1-S5) removed. Although we are concerned with the type of fault acting on a single qubit, qubits interact with each other by means of the tensor product such that the dual qubit representation for faults acting on Q1 follows,

| $\begin{bmatrix} 0 & 0 & 1 & 0 \\ 0 & 0 & 0 & 1 \\ 1 & 0 & 0 & 0 \\ 0 & 1 & 0 & 0 \end{bmatrix}$ | $\begin{bmatrix} 0 & 0 & -j & 0 \\ 0 & 0 & 0 & -j \\ j & 0 & 0 & 0 \\ 0 & j & 0 & 0 \end{bmatrix}$ | $\begin{bmatrix} -1 & 0 & 0 & 0 \\ 0 & -1 & 0 & 0 \\ 0 & 0 & 1 & 0 \\ 0 & 0 & 0 & 1 \end{bmatrix}$ |
|---|---|---|
| Pauli – X@Q1 | Pauli – Y@Q1 | Pauli – Z@Q1 |

And for the faults acting on the second qubit (Q2) we have,

| $\begin{bmatrix} 0 & 1 & 0 & 0 \\ 1 & 0 & 0 & 0 \\ 0 & 0 & 0 & 1 \\ 0 & 0 & 1 & 0 \end{bmatrix}$ | $\begin{bmatrix} 0 & -j & 0 & 0 \\ j & 0 & 0 & -j \\ 0 & 0 & j & 0 \\ 0 & 0 & 0 & 0 \end{bmatrix}$ | $\begin{bmatrix} -1 & 0 & 0 & 0 \\ 0 & 1 & 0 & 0 \\ 0 & 0 & -1 & 0 \\ 0 & 0 & 0 & 1 \end{bmatrix}$ |
|---|---|---|
| Pauli-X@Q2 | Pauli-Y@Q2 | Pauli-Z@Q2 |

The basic steps of the presented algorithm start with the calculation of the expected linear operator that represents how this sequence of gates will transform the state of a qubit. From Equation 2 we have shown that this pulse sequence is a CNOT gate relative to phase,

$$GC = \frac{(1-j)}{\sqrt{2}} \cdot \begin{bmatrix} 1 & 0 & 0 & 0 \\ 0 & 0 & 0 & 1 \\ 0 & 0 & 1 & 0 \\ 0 & 1 & 0 & 0 \end{bmatrix} \Leftrightarrow \begin{array}{cc|cc} a & b & a & b' \\ \hline 0 & 0 & 0 & 0 \\ 0 & 1 & 1 & 1 \\ 1 & 0 & 1 & 0 \\ 1 & 1 & 0 & 1 \end{array}$$

*Figure 8 – Phase relative linear operator representing the CNOT gate, and its corresponding truth table.*

As the first case we will consider the Pauli-X fault in a location on the first qubit (Q1). For the circuit under observation this is meant to reside in a total of six possible locations. The Pauli-X fault is inserted independently into those locations and the linear transformation performed by the erroneous circuit is again recalculated using QuIDDPro [1] and placed in the following table,

| $\frac{(1-j)}{\sqrt{2}} \cdot \begin{bmatrix} 0 & 0 & 1 & 0 \\ 0 & 1 & 0 & 0 \\ 1 & 0 & 0 & 0 \\ 0 & 0 & 0 & 1 \end{bmatrix}$ | $\frac{(1-j)}{\sqrt{2}} \cdot \begin{bmatrix} 0 & 0 & 1 & 0 \\ 0 & 1 & 0 & 0 \\ 1 & 0 & 0 & 0 \\ 0 & 0 & 0 & 1 \end{bmatrix}$ | $\frac{(1-j)}{\sqrt{2}} \cdot \begin{bmatrix} 0 & 0 & 1 & 0 \\ 0 & 1 & 0 & 0 \\ 1 & 0 & 0 & 0 \\ 0 & 0 & 0 & 1 \end{bmatrix}$ |
|---|---|---|
| X – Q1@D1 | X – Q1@D2 | X – Q1@D3 |
| $\frac{(1+j)}{\sqrt{2}} \cdot \begin{bmatrix} 0 & 0 & -1 & 0 \\ 0 & -1 & 0 & 0 \\ 1 & 0 & 0 & 0 \\ 0 & 0 & 0 & 1 \end{bmatrix}$ | $\frac{(1-j)}{\sqrt{2}} \cdot \begin{bmatrix} -1 & 0 & 0 & 0 \\ 0 & 0 & 0 & 1 \\ 0 & 0 & 1 & 0 \\ 0 & -1 & 0 & 0 \end{bmatrix}$ | $\frac{(1-j)}{\sqrt{2}} \cdot \begin{bmatrix} 0 & 0 & 1 & 0 \\ 0 & 1 & 0 & 0 \\ 1 & 0 & 0 & 0 \\ 0 & 0 & 0 & 1 \end{bmatrix}$ |
| X – Q1@D4 | X – Q1@D5 | X – Q1@D6 |

from these results the reversible fault table immediately follows,

| Inputs a b | GC | X – Q1@D1 | X – Q1@D2 | X – Q1@D3 | X – Q1@D4 | X – Q1@D5 | X – Q1@D6 |
|---|---|---|---|---|---|---|---|
| 0 0 | 0 0 | **1 0** | **1 0** | **1 0** | **1 0** | 0 0 | **1 0** |
| 0 1 | 1 1 | **0 1** | **0 1** | **0 1** | **0 1** | 1 1 | **0 1** |
| 1 0 | 1 0 | **0 0** | **0 0** | **0 0** | **0 0** | 1 0 | **0 0** |
| 1 1 | 0 1 | **1 1** | **1 1** | **1 1** | **1 1** | 0 1 | **1 1** |

*(Reversible truth table from the insertion of Pauli-X at locations D1-D6 on the first qubit from Figure 6)*

We note that for the Pauli-X fault impacting Q1 that all test inputs are equivalent for this case except the apparent lack of measured error with Pauli-X impacting Q1 at D5, we will not address this. It is the typical case for certain technologies such as NMR that one can measure the circuit before or after the expected fault to gain insight into its presence, we do not address this method as we deal here only with detecting faults in the logical outputs of permutative circuits, from the computational basis. In a similar manor to that already done for the single Pauli-X fault on the first qubit we repeat the process for the Pauli-Y and Pauli-Z rotations respectively,

| $\frac{(1+j)}{\sqrt{2}} \cdot \begin{bmatrix} 0 & 0 & -1 & 0 \\ 0 & -1 & 0 & 0 \\ 1 & 0 & 0 & 0 \\ 0 & 0 & 0 & 1 \end{bmatrix}$ | $\frac{(1+j)}{\sqrt{2}} \cdot \begin{bmatrix} 0 & 0 & -1 & 0 \\ 0 & -1 & 0 & 0 \\ 1 & 0 & 0 & 0 \\ 0 & 0 & 0 & 1 \end{bmatrix}$ | $\frac{(1-j)}{\sqrt{2}} \cdot \begin{bmatrix} 1 & 0 & 0 & 0 \\ 0 & 0 & 0 & 1 \\ 0 & 0 & -1 & 0 \\ 0 & -1 & 0 & 0 \end{bmatrix}$ |
|---|---|---|
| Y – Q1@D1 | Y – Q1@D2 | Y – Q1@D3 |
| $\frac{(1-j)}{\sqrt{2}} \cdot \begin{bmatrix} 1 & 0 & 0 & 0 \\ 0 & 0 & 0 & 1 \\ 0 & 0 & -1 & 0 \\ 0 & -1 & 0 & 0 \end{bmatrix}$ | $\frac{(1+j)}{\sqrt{2}} \cdot \begin{bmatrix} 0 & 0 & -1 & 0 \\ 0 & 1 & 0 & 0 \\ 1 & 0 & 0 & 0 \\ 0 & 0 & 0 & -1 \end{bmatrix}$ | $\frac{(1+j)}{\sqrt{2}} \cdot \begin{bmatrix} 0 & 0 & -1 & 0 \\ 0 & 1 & 0 & 0 \\ 1 & 0 & 0 & 0 \\ 0 & 0 & 0 & -1 \end{bmatrix}$ |
| Y – Q1@D4 | Y – Q1@D5 | Y – Q1@D6 |

*(Transformations of the circuit from Figure 7 with the addition of the Pauli-Y fault in the labeled locations)*

| Inputs a b | GC | Y – Q1@D1 | Y – Q1@D2 | Y – Q1@D3 | Y – Q1@D4 | Y – Q1@D5 | Y – Q1@D6 |
|---|---|---|---|---|---|---|---|
| 0 0 | 0 0 | **1 0** | **1 0** | 0 0 | 0 0 | **1 0** | **1 0** |
| 0 1 | 1 1 | **0 1** | **0 1** | 1 1 | 1 1 | **0 1** | **0 1** |
| 1 0 | 1 0 | **0 0** | **0 0** | 1 0 | 1 0 | **0 0** | **0 0** |
| 1 1 | 0 1 | **1 1** | **1 1** | 0 1 | 0 1 | **1 1** | **1 1** |

*(The reversible fault table created by analyzing the possible outcomes of measurement of the Pauli-Y fault acting on Q1)*

We note the equivalence of inputs and tests for a Pauli-Y fault impacting the first Qubit at locations D1, D2, D5, and D6. However, a similar fault at location D4 or D5 remains undetectable if only logical output is compared from our current measurement basis.

| $\frac{(1-j)}{\sqrt{2}} \cdot \begin{bmatrix} -1 & 0 & 0 & 0 \\ 0 & 0 & 0 & -1 \\ 0 & 0 & 1 & 0 \\ 0 & 1 & 0 & 0 \end{bmatrix}$ | $\frac{(1-j)}{\sqrt{2}} \cdot \begin{bmatrix} -1 & 0 & 0 & 0 \\ 0 & 0 & 0 & -1 \\ 0 & 0 & 1 & 0 \\ 0 & 1 & 0 & 0 \end{bmatrix}$ | $\frac{(1+j)}{\sqrt{2}} \cdot \begin{bmatrix} 0 & 0 & -1 & 0 \\ 0 & -1 & 0 & 0 \\ 1 & 0 & 0 & 0 \\ 0 & 0 & 0 & 1 \end{bmatrix}$ |
|---|---|---|
| Z – Q1@D1 | Z – Q1@D2 | Z – Q1@D3 |
| $\frac{(j-1)}{\sqrt{2}} \cdot \begin{bmatrix} 0 & 0 & 1 & 0 \\ 0 & 1 & 0 & 0 \\ 1 & 0 & 0 & 0 \\ 0 & 0 & 0 & 1 \end{bmatrix}$ | $\frac{(1+j)}{\sqrt{2}} \cdot \begin{bmatrix} 0 & 0 & -1 & 0 \\ 0 & -1 & 0 & 0 \\ -1 & 0 & 0 & 0 \\ 0 & 0 & 0 & -1 \end{bmatrix}$ | $\frac{(1-j)}{\sqrt{2}} \cdot \begin{bmatrix} -1 & 0 & 0 & 0 \\ 0 & 0 & 0 & 1 \\ 0 & 0 & 1 & 0 \\ 0 & -1 & 0 & 0 \end{bmatrix}$ |
| Z – Q1@D4 | Z – Q1@D5 | Z – Q1@D6 |

*(Transformations of the circuit shown with the addition of the Pauli-Z fault in the labeled locations all acting on Q1)*

As can be seen from Table 6 the impact of the Pauli-Z fault at locations D1, D2, and D6 are not detectable from the current basis. For the same type of fault at locations D3, D4, and D5 the fault is both detectable and all input tests are equivalent, greatly simplifying the testing process for those cases.

| Inputs: a b | GC | Z – Q1@D1 | Z – Q1@D2 | Z – Q1@D3 | Z – Q1@D4 | Z – Q1@D5 | Z – Q1@D6 |
|---|---|---|---|---|---|---|---|
| 0 0 | 0 0 | 0 0 | 0 0 | **1 0** | **1 0** | **1 0** | 0 0 |
| 0 1 | 1 1 | 1 1 | 1 1 | **0 1** | **0 1** | **0 1** | 1 1 |
| 1 0 | 1 0 | 1 0 | 1 0 | **0 0** | **0 0** | **0 0** | 1 0 |
| 1 1 | 0 1 | 0 1 | 0 1 | **1 1** | **1 1** | **1 1** | 0 1 |

*Table 6 – Pauli-Z reversible Fault Table*

We now show the results of repeating this entire process of finding the linear operator representing faults acting on Q1 for Q2, starting with the Pauli-X on Q2,

| $\frac{(1-j)}{\sqrt{2}} \cdot \begin{bmatrix} 0 & 0 & 0 & 1 \\ 1 & 0 & 0 & 0 \\ 0 & 1 & 0 & 0 \\ 0 & 0 & 1 & 0 \end{bmatrix}$ | $\frac{(1+j)}{\sqrt{2}} \cdot \begin{bmatrix} 0 & 0 & 0 & -1 \\ 1 & 0 & 0 & 0 \\ 0 & -1 & 0 & 0 \\ 0 & 0 & 1 & 0 \end{bmatrix}$ | $\frac{(1+j)}{\sqrt{2}} \cdot \begin{bmatrix} 0 & 0 & 0 & -1 \\ 1 & 0 & 0 & 0 \\ 0 & -1 & 0 & 0 \\ 0 & 0 & 1 & 0 \end{bmatrix}$ |
|---|---|---|
| X – Q2@D1 | X – Q2@D2 | X – Q2@D3 |
| $\frac{(1+j)}{\sqrt{2}} \cdot \begin{bmatrix} 0 & 0 & 0 & -1 \\ 1 & 0 & 0 & 0 \\ 0 & -1 & 0 & 0 \\ 0 & 0 & 1 & 0 \end{bmatrix}$ | $\frac{(1-j)}{\sqrt{2}} \cdot \begin{bmatrix} 0 & 1 & 0 & 0 \\ 0 & 0 & 1 & 0 \\ 0 & 0 & 0 & 1 \\ 1 & 0 & 0 & 0 \end{bmatrix}$ | $\frac{(1-j)}{\sqrt{2}} \cdot \begin{bmatrix} 0 & 1 & 0 & 0 \\ 0 & 0 & 1 & 0 \\ 0 & 0 & 0 & 1 \\ 1 & 0 & 0 & 0 \end{bmatrix}$ |
| X – Q2@D4 | X – Q2@D5 | X – Q2@D6 |

*Resulting linear operators from insertion of the Pauli-X fault at the labeled locations in Figure 7*

| Inputs a b | GC | X – Q2@D1 | X – Q2@D2 | X – Q2@D3 | X – Q2@D4 | X – Q2@D5 | X – Q2@D6 |
|---|---|---|---|---|---|---|---|
| 0 0 | 0 0 | **0 1** | **0 1** | **0 1** | **0 1** | **1 1** | **1 1** |
| 0 1 | 1 1 | **1 0** | **1 0** | **1 0** | **1 0** | **0 0** | **0 0** |
| 1 0 | 1 0 | **1 1** | **1 1** | **1 1** | **1 1** | **0 1** | **0 1** |
| 1 1 | 0 1 | **0 0** | **00** | **00** | **00** | **1 0** | **1 0** |

*(The reversible truth table from the Pauli-X fault acting on Q2)*

We note that for the Pauli-X fault impacting Q2 that all test inputs are equivalent in detection of error. And next we calculate the erroneous linear transformation resulting from the Pauli-Y fault acting on Q2, from this we have,

| $\frac{(1+j)}{\sqrt{2}} \cdot \begin{bmatrix} 0 & 0 & 0 & -1 \\ 1 & 0 & 0 & 0 \\ 0 & -1 & 0 & 0 \\ 0 & 0 & 1 & 0 \end{bmatrix}$ | $\frac{(1-j)}{\sqrt{2}} \cdot \begin{bmatrix} 0 & 0 & 0 & -1 \\ -1 & 0 & 0 & 0 \\ 0 & -1 & 0 & 0 \\ 0 & 0 & -1 & 0 \end{bmatrix}$ | $\frac{(1-j)}{\sqrt{2}} \cdot \begin{bmatrix} 0 & 0 & 0 & -1 \\ -1 & 0 & 0 & 0 \\ 0 & -1 & 0 & 0 \\ 0 & 0 & -1 & 0 \end{bmatrix}$ |
|---|---|---|
| Y – Q2@D1 | Y – Q2@D2 | Y – Q2@D3 |
| $\frac{(1-j)}{\sqrt{2}} \cdot \begin{bmatrix} 0 & 0 & 0 & -1 \\ -1 & 0 & 0 & 0 \\ 0 & -1 & 0 & 0 \\ 0 & 0 & -1 & 0 \end{bmatrix}$ | $\frac{(1+j)}{\sqrt{2}} \cdot \begin{bmatrix} 0 & -1 & 0 & 0 \\ 0 & 0 & 1 & 0 \\ 0 & 0 & 0 & -1 \\ 1 & 0 & 0 & 0 \end{bmatrix}$ | $\frac{(1+j)}{\sqrt{2}} \cdot \begin{bmatrix} 0 & -1 & 0 & 0 \\ 0 & 0 & 1 & 0 \\ 0 & 0 & 0 & -1 \\ 1 & 0 & 0 & 0 \end{bmatrix}$ |
| Y – Q2@D4 | Y – Q2@D5 | Y – Q2@D6 |

*(Resulting linear operators from insertion of the Pauli-Y fault at the labeled locations in Figure 7)*

| Inputs a b | GC | Y – Q2@D1 | Y – Q2@D2 | Y – Q2@D3 | Y – Q2@D4 | Y – Q2@D5 | Y – Q2@D6 |
|---|---|---|---|---|---|---|---|
| 0 0 | 0 0 | **0 1** | **0 1** | **0 1** | **0 1** | **1 1** | **1 1** |
| 0 1 | 1 1 | **1 0** | **1 0** | **1 0** | **1 0** | **0 0** | **0 0** |
| 1 0 | 1 0 | **1 1** | **1 1** | **1 1** | **1 1** | **0 1** | **0 1** |
| 1 1 | 0 1 | **0 0** | **0 0** | **0 0** | **0 0** | **1 0** | **1 0** |

*(The reversible truth table from the Pauli-Y fault acting on Q2)*

We note that in the case of the Pauli-Y fault impacting Q2 that all test vectors are equivalent. And for Pauli-Z acting on the second Qubit (Q2) we have,

| $\frac{(1-j)}{\sqrt{2}} \cdot \begin{bmatrix} -1 & 0 & 0 & 0 \\ 0 & 0 & 0 & 1 \\ 0 & 0 & -1 & 0 \\ 0 & 1 & 0 & 0 \end{bmatrix}$ | $\frac{(1-j)}{\sqrt{2}} \cdot \begin{bmatrix} -1 & 0 & 0 & 0 \\ 0 & 0 & 0 & 1 \\ 0 & 0 & -1 & 0 \\ 0 & 1 & 0 & 0 \end{bmatrix}$ | $\frac{(1-j)}{\sqrt{2}} \cdot \begin{bmatrix} -1 & 0 & 0 & 0 \\ 0 & 0 & 0 & 1 \\ 0 & 0 & -1 & 0 \\ 0 & 1 & 0 & 0 \end{bmatrix}$ |
|---|---|---|
| Z – Q2@D1 | Z – Q2@D2 | Z – Q2@D3 |

| | $\frac{(1-j)}{\sqrt{2}} \cdot \begin{bmatrix} -1 & 0 & 0 & 0 \\ 0 & 0 & 0 & 1 \\ 0 & 0 & -1 & 0 \\ 0 & 1 & 0 & 0 \end{bmatrix}$ | $\frac{(1-j)}{\sqrt{2}} \cdot \begin{bmatrix} -1 & 0 & 0 & 0 \\ 0 & 0 & 0 & 1 \\ 0 & 0 & -1 & 0 \\ 0 & 1 & 0 & 0 \end{bmatrix}$ | $\frac{(1-j)}{\sqrt{2}} \cdot \begin{bmatrix} -1 & 0 & 0 & 0 \\ 0 & 0 & 0 & 1 \\ 0 & 0 & -1 & 0 \\ 0 & 1 & 0 & 0 \end{bmatrix}$ |
|---|---|---|---|
| | Z – Q2@D4 | Z – Q2@D5 | Z – Q2@D6 |

*(Resulting linear operators from insertion of the Pauli-Y fault at the labeled locations in Figure 7)*

As shown in the following table the Pauli-Z fault does not appear to impact the outcome of measurement if acting on Q2 from the computational basis as provided in the table below.

| Inputs a b | GC | Z – Q2@D1 | Z – Q2@D2 | Z – Q2@D3 | Z – Q2@D4 | Z – Q2@D5 | Z – Q2@D6 |
|---|---|---|---|---|---|---|---|
| 0 0 | 0 0 | 0 0 | 0 0 | 0 0 | 0 0 | 0 0 | 0 0 |
| 0 1 | 1 1 | 1 1 | 1 1 | 1 1 | 1 1 | 1 1 | 1 1 |
| 1 0 | 1 0 | 1 0 | 1 0 | 1 0 | 1 0 | 1 0 | 1 0 |
| 1 1 | 0 1 | 0 1 | 0 1 | 0 1 | 0 1 | 0 1 | 0 1 |

*(The reversible truth table from the Pauli-Z fault acting on Q2)*

Now we will replace each gate in the sequence with that of the corresponding identity (pulses S1-S5),

| $\begin{bmatrix} 1 & 0 & 0 & 0 \\ 0 & 0 & 0 & -j \\ 0 & 0 & 1 & 0 \\ 0 & -j & 0 & 0 \end{bmatrix}$ | $\frac{1}{2} \cdot \begin{bmatrix} 1+j & 0 & 1+j & 0 \\ 0 & 1+j & 0 & 1-j \\ 1+j & 0 & 1-j & 0 \\ 0 & 1-j & 0 & 1+j \end{bmatrix}$ | $\begin{bmatrix} 1 & 0 & 0 & 0 \\ 0 & 0 & 0 & -j \\ 0 & 0 & 1 & 0 \\ 0 & -j & 0 & 0 \end{bmatrix}$ |
|---|---|---|
| S1 – removed | S2 – removed | S3 – removed |
| $\begin{bmatrix} -j & 0 & 0 & 0 \\ 0 & j & 0 & 0 \\ 0 & 0 & -j & 0 \\ 0 & 0 & 0 & j \end{bmatrix}$ | $\frac{(1-j)}{2} \cdot \begin{bmatrix} 1 & 0 & -1 & 0 \\ 0 & 1 & 0 & 1 \\ 1 & 0 & 1 & 0 \\ 0 & 1 & 0 & -1 \end{bmatrix}$ | |
| S4 – removed | S5 – removed | |

*(Resulting linear operators from removal of stages S1-S5 from Figure 7)*

The removal of stages S1-S5 leads one to create that defined here as a quantum fault table. As mentioned we distinguish a classical fault table and a quantum fault table with the later having the addition of positive fractions and the former restrained to the binary set. In the table below the prescribed expected measurement predicted occurrence rate is noted as a percentage for those non 1 to the right of the binary outcome.

| Inputs a b | GC | Section Removed | | | | | |
|---|---|---|---|---|---|---|---|
| | | S1 | S2 | S3 | S4 | S5 | |
| 0 0 | 0 0 | 0 0 | 0(50%)<br>**1(50%)** | 0 | 0 0 | 0 0 | 0(50%)<br>**1(50%)** | 0 |
| 0 1 | 1 1 | 1 1 | **0(50%)**<br>1(50%) | 1 | 1 1 | **0 1** | 0(50%)<br>**1(50%)** | 1 |
| 1 0 | 1 0 | 1 0 | **0(50%)**<br>1(50%) | 0 | 1 0 | 1 0 | 0(50%)<br>**1(50%)** | 0 |
| 1 1 | 0 1 | 0 1 | 0(50%)<br>**1(50%)** | 1 | 0 1 | **1 1** | 0(50%)<br>**1(50%)** | 1 |

*(The reversible truth table from individually removing stages S1 through S5 from Figure 7)*

As an example we will now construct a non reversible quantum fault table,

| Inputs a b | Section Removed | | | | |
|---|---|---|---|---|---|
| | S1 | S2 | S3 | S4 | S5 |
| 0 0 | 0 | **.5** | 0 | 0 | **.5** |
| 0 1 | 0 | .5 | 0 | **1** | .5 |
| 1 0 | 0 | .5 | 0 | 0 | .5 |
| 1 1 | 0 | **.5** | 0 | **1** | **.5** |

⇔

| Inputs a b | Section Removed | |
|---|---|---|
| | S2, S5 | S4 |
| 0 0 | **.5** | 0 |
| 0 1 | .5 | **1** |
| 1 0 | .5 | 0 |
| 1 1 | **.5** | **1** |

*(The simplification of the reversible truth table built from individually removing stages S1 through S5 from Figure 7)*

| Test Vector | Pauli-X acting on Q1 (D1, D2, D3, D4, D6) | Pauli-Y acting on Q1 (D1, D2, D5, D6) | Pauli-Z acting on Q1 (D3, D4, D5, D6) | Pauli-X acting on Q2 (D1, D2, D3, D4, D5, D6) | Pauli-Y acting on Q2 (D1, D2, D3, D4, D5, D6) | Removal of section: S2, S5 | Removal of section: S4 |
|---|---|---|---|---|---|---|---|
| T1(00) | 1 | 1 | 1 | 1 | 1 | **.5** | 0 |
| T2(01) | 1 | 1 | 1 | 1 | 1 | .5 | **1** |
| T3(10) | 1 | 1 | 1 | 1 | 1 | .5 | 0 |
| T4(11) | 1 | 1 | 1 | 1 | 1 | **.5** | **1** |

*Table 7 – Quantum Fault Table built from Figure 7*

It is worth pointing out that this table is built by superimposing entries that are the same for each of the fault tables for each respective fault for both bits, and removing entries representing faults that had no impact on the circuit from Figure 7. In this example the problem of quantum testing reduces to only partial cover for the detection of the removal of S2 and S5. It is the goal of future work to analyze the behavior of many different circuits of varying sizes [38], in doing such one is made to encounter tables that are much more complicated than the example just presented. In general, a Quantum Fault Tables solution requires a solution to the fractional covering problem. This formulation will allow testing for any fault model inserted, given the restraints of our problem scope. From Table 7 we will select T4 as the first test to execute leaving only S2 and S5 possibly undetected, from the computational basis. S2 and S5 can be still be covered to a certain value using iterative testing.

## 9. CONCLUSIONS AND FUTURE WORK

We showed a method to minimize the number of tests needed to verify the correct operation of a quantum circuit. The method based on the Quantum Fault Table is general and does not depend on any particular fault model, it assumes however the existence of a single fault in the quantum circuit. If more faults exist, the method should still detect them with high probability, but this probability may be lower than that evaluated by this method. This work has opened the door to what is sure to be an extremely useful area of research, as the experimental physicist will soon need verification methods better than the current testing methods that take exponentially more time with circuit size. Further studies of fault models and test generation designed for fault diagnostics are the authors' current research topic.

## 10. ACKNOWLEDGEMENTS

We are grateful for the excellent facilities of the Korean Advanced Institute of Science and Technology, where some of our research was conducted. We are also grateful to Professor Soonchill Lee of the Korean Advanced Institute of Science and Technology, George F. Viamontes the developer of QuIDDPro [1], and Dr. Alan Mishchenko all who added valuable comments suggestions and or materials that aided in the creation of this paper. Although they helped in the creation of this work, the remaining errors are of course our own. This work was supported by Ronald E. McNair Post baccalaureate Achievement Program of Portland State University, Portland State University, and the Korean Institute of Science and Technology. The views and conclusions contained herein are those of the authors and should not be interpreted as necessarily representing official policies of endorsements, either expressed or implied, of the Funding agencies. ◊

# WORK CITED

1. G. F. Viamontes, I. L. Markov, and J. P. Hayes, Improving gate-level simulation of quantum circuits, *Quantum Information Processing, vol. 2(5), pp. 347-380, October 2003* http://xxx.lanl.gov/abs/quant-ph/0309060
2. K. N. Patel, J. P. Hayes and I. L. Markov, Fault Testing for Reversible Circuits, quant-ph/0404003
3. R.P. Feynman: There's plenty of room at the bottom, *printed in February 1960, Caltech's Engineering Science*
4. G. E. Moore, Cramming more components onto integrated circuits, *Electronics, Volume 38, Number 8, April 19, 1965*
5. R. Landauer, Irreversibility and heat generation in the computing process, *IBM J. Res. Dev., 5 p.183, 1961*
6. C.H. Bennett, Logical reversibility of computation, *IBM J. Res. Dev. 17 p.525 1973*
7. R.P. Feynman: Simulating physics with computers, *Int. J. of Theor. Phys., 21, 1982*
8. L. M.K. Vandersypen, M. Steffen, G. Breyta, C. S. Yannoni, M. H. Sherwood, I. L. Chuang, Experimental realization of Shor's quantum factoring algorithm using nuclear magnetic resonance, *Nature 414, 883-887,20/27, Dec 2001*
9. P. W. Shor, Polynomial-time algorithms for prime factorization and discreate logrythems on a quantum computer, *SIAM J. Comp., 26(5): 1484-1509, 1997*
10. L. K. Grover, Quantum Mechanics helps in searching for a needle in a haystack, *Phys. Rev. Lett., 79(2):325, 1997.* arXive-print quant-ph/9706033
11. J. Ankerhold and H. Grabert, Enhancement of Macroscopic Quantum Tunneling by Landau–Zener Transitions, *Science, volume 296, page 886, May 2002*
12. D. Vion, A. Aassime, A. Cottet, P. Joyez, H. Pothier, C. Urbina, D. Esteve, M.H. Devoret, Manipulating the Quantum State of an Electrical Circuit, *arXiv:cond-mat/0304232 v2 15 Apr 2003*
13. H-J. Wunderlich, S. Hellebrand, The Pseudo-Exhaustive Test of Sequential Circuits, *Proceedings IEEE International Test Conference, Washington, DC, 1989*
14. C. Landrault, Translated by M. A. Perkowski, Test and Design For Test, *www.ee.pdx.edu/~mperkows*
15. J. Biamonte, M. Perkowski, Principles of Quantum Fault Detection, *Portland State University INQ research conference, June 08, 2004*
16. J. Biamonte, M. Perkowski, Principles of Quantum Fault Diagnostics, *to appear in McNair research Journal,* Issue 1, Volume 1, 2004
17. S. Reddy, Easily Testable Realizations for Logic Functions, 1972
18. A.M. Brosa and J. Figueras, Digital Signature Proposal for Mixed-Signal Circuits
19. E. McCluskey and Ch-W. Tseng, Stuck-Fault Tests vs. Actual Defects, 1997
20. M. Nielsen, I. Chuang, Quantum Computing and Quantum Information, *Cambridge University Press, 2000*
21. C. Williams, S. Clearwater, Explorations in Quantum Computing, *Springer Press, 1997*
22. Ashikhmin, A. Barg, E. Knill, and S. Litsyn, Quantum error detection I: Statement of the problem, *IEEE Trans. Inf. Theory, 1999.* arXiv, quant-ph/9906126
23. Ashikhmin, A. Barg, E. Knill, and S. Litsyn, Quantum error detection II: Bounds. *IEEE Trans. Inf. Theory, 1999.* arXiv, quant-ph/9906131
24. H.Bennett, D. P. DiVincenzo, J. A. Smolin, and W. K. Wootters, Mixed state entanglement and quantum error-correcting codes, *Phys. Rev. A, 54:3824, 1996.* arXiv, quant-ph/9604024
25. T. Sasao, Easily Testable Realizations for Generalized Reed-Muller Expressions, IEEE Transactions on Computers, Vol. 46, No. 6, June 1997, pp. 709-716.
26. J. Scott, Probabilities of failure for quantum error correction, *arXiv:quant-ph/0406063 v1 9 Jun 2004*
27. S. Bettelli, Quantitative model for the effective decoherence of a quantum computer with imperfect unitary operations, *Physical Review A 69,042310, 14 pages, 2004*
28. Deutsch, The Church-Turing principle and the universal quantum computer, *Proceedings of the Royal Society of London A 400, pp. 97-117, 1985*
29. S. Lee, S-J. Lee, T. Kim, J-S. Lee, J. Biamonte and M. Perkowski, The Cost of Quantum Gate Primitives, *submitted to the International Journal of Multi-valued Logic and Soft Computing, 2004*
30. J. Kim, J-S. Lee, S. Lee, Implementing unitary operators in quantum computation, quant-ph/9908052
31. R. C. Merkle, Reversible electronic logic using switches, *Nanotechnology, 4: pp. 21-40, 1993*
32. R. C. Merkle, Two types of mechanical reversible logic, *Nanotechnology, 4: pp. 114-131,1993*
33. Fredkin, T. Toffoli, Conservative Logic, *MIT Laboratory for Computer Science 545 Technology Square, Cambridge, Massachusetts 02139*
34. W. Zurek, Reversibility and Stability of Information Processing Systems, *Physical Review Letters, Vol. 53, pp. 391-394, 1984*
35. S. Kak, The Initialization Problem in Quantum Computing, *Foundations of Physics, vol 29, pp. 267-279, 1999*
36. K. Ramasamy, R. Tagare, E. Perkins and M. Perkowski, Fault localization in reversible circuits is easier than for classical circuits, *Proceedings of the International Workshop on Logic and Synthesis, June 2004.*
37. P. Beraldi, A. Ruszczy, The Probabilistic Set Sovering Problem, *Operations Research © 2002 INFORMS, Vol.50, No.6, November–December 2002, pp. 956–967*
38. J. Biamonte, Principles of Quantum Test, *to be submitted 2004.*